\documentclass[12pt]{article}
\usepackage{epsfig,wrapfig,amsmath}
\renewcommand{\theequation}{\arabic{section}.\arabic{equation}}

\def\beqn{\begin{eqnarray}}
\def\eeqn{\end{eqnarray}}

\begin{document}

\thispagestyle{empty}
\renewcommand{\thefootnote}{\fnsymbol{footnote}}
\begin{flushleft}
\large {SAGA-HE-129-98  \hfill 
January 10, 1998} \\
\end{flushleft}
 
\vspace{2.0cm}
 
\begin{center}
{\Large \bf Studies of polarized parton distributions at RHIC \\ }

\vspace{1.8cm}

\Large
S. Kumano $^*$

\vspace{1.0cm}
 
Department of Physics  \\
 
\vspace{0.1cm}
 
Saga University \\
 
\vspace{0.1cm}

Saga 840-8502, Japan \\

\vspace{1.5cm}

\large 
Talk given at the Symposium on ``RHIC Physics"\\

\vspace{0.3cm}

Japanese Physical Society Meeting \\

\vspace{0.7cm}

Tokyo, Japan, Sept. 20 $-$ 23, 1997 \\
(talk on Sept. 21, 1997) \\
 
\end{center}

\vspace{1.2cm}

\vfill
 
\noindent
{\rule{6.cm}{0.1mm}} \\
 
\vspace{-0.2cm}
\normalsize
\noindent
{* Email: kumanos@cc.saga-u.ac.jp.} \\

\vspace{-0.6cm}
{Information on his research is available at}  \\

\vspace{-0.6cm}
{http://www.cc.saga-u.ac.jp/saga-u/riko/physics/quantum1/structure.html.} \\

\vspace{+0.5cm}
\hfill
{to be published in Genshikaku Kenkyu}

\clearpage
\setcounter{page}{1}

$\ \ \ $

$\ \ \ $

$\ \ \ $

$\ \ \ $

$\ \ \ $

\begin{center}
{\Large \bf Studies of polarized parton distributions at RHIC \\ }
\vspace{5mm}

S. Kumano $^*$

\vspace{5mm}
{\small\it
Department of Physics, Saga University, Saga 840-8502, Japan
\\ }

\vspace{5mm}
ABSTRACT

\vspace{5mm}
\begin{minipage}{130 mm}
\small


We discuss our studies of polarized parton distributions which are related
to the RHIC-Spin project. First, the parametrization of unpolarized parton
distributions is explained as an introduction to general audience.
Second, activities of the RHIC-Spin-J working group are reported. 
In particular, preliminary results are shown on the parametrization of
polarized parton distributions. Third, we discuss numerical solution
of the DGLAP $Q^2$ evolution equations for longitudinally
polarized and transversity distributions.
An efficient and accurate computer program is important for analyzing
future RHIC experimental data.
 
\end{minipage}
\end{center}

\section{{\bf Introduction}}\label{INTRO}
\setcounter{equation}{0}
\setcounter{figure}{0}
\setcounter{table}{0}

Internal spin structure of the proton has been investigated
extensively for the last ten years. 
Although the proton spin is no more considered at the stage of
``crisis", the clear explanation is still far from the final one.
In particular, there is essentially no information on
sea-quark and gluon polarizations. Furthermore, parton's angular
momenta may have influence on the issue. In order to clarify
these problems, the RHIC-Spin project was proposed \cite{RHIC-SPIN}.
Obtained data at the RHIC facility should be able to pin down
how the proton spin consists of internal constituent spins.

The RHIC facility will provide us fruitful information
on the proton spin structure. In order to understand the situation
of proton-spin studies and also to explore possible RHIC-Spin physics,
we established a Japanese working group on parametrization of polarized
parton distributions \cite{RHIC-SPIN-J}. For the time being, the purpose
is to obtain the optimum parton distributions for explaining existing
polarized experimental data. Our study is still at the 
preliminary stage \cite{JPS}.
The parametrization of unpolarized parton distributions has been
well studied since 1980's. It is also fortunate that abundant
experimental data are available through various reactions
in the unpolarized case. Accurate distributions are now known
from very small $x$ to relatively large $x$. 
The same parametrization study can be done for the longitudinally
polarized parton distributions because many experimental data are now
available for the structure function $g_1$ \cite{PARA}.

The proton spin is so far investigated mainly in the longitudinally
polarized structure function $g_1$. However, it is also interesting
to test it by transversely polarized structure functions.
In particular, the leading-twist structure function $h_1$ should be
an interesting topic. It is expected to be measured in the
transversely polarized Drell-Yan processes at RHIC. 
We should try to understand the properties of $h_1$ before
the experimental data are taken. 
The transversity distribution is an unexplored subject experimentally.
If future experimental data on $h_1$ are much different from
theoretical expectation, a new field of spin physics could be developed.

We have been working on calculation of
the next-to-leading-order (NLO) anomalous dimensions
for $h_1$ and also on numerical solution of its evolution equation
\cite{T-EVOL}. We also studied the numerical solution
of the Dokshitzer-Gribov-Lipatov-Altarelli-Parisi (DGLAP)
evolution equations for the longitudinally polarized 
distributions \cite{L-EVOL}. In general, parton distributions are
provided at $Q^2 \sim O$(1 GeV$^2$) and experimental data are taken
at different $Q^2$ points. Therefore, our studies on the numerical
solution should be useful for extracting information
on the polarized parton distributions from polarized RHIC data.

The major audience are traditional nuclear physicists at this symposium,
so that we first discuss how each unpolarized parton distribution
is obtained from various experimental data in section \ref{EXTRACT}.
The section \ref{PARAMET} is devoted to the activities of the RHIC-Spin-J
working group. A brief outline of the parametrization project is explained.
We discuss the numerical solution of the DGLAP $Q^2$ evolution equations
for the longitudinally polarized and transversity distributions 
in section \ref{EVOL}. The summary is given in section \ref{SUM}.

\section{{\bf Extraction of parton distributions}}\label{EXTRACT}
\setcounter{equation}{0}

Nucleon substructure has been investigated through various high-energy
experiments. For example, electron or muon deep inelastic scattering
has been used. Its unpolarized cross section is related to two structure
functions $F_1$ and $F_2$. They depend in general on two kinematical
variables $Q^2=-q^2$ and $x=Q^2/2p\cdot q$ where $q$ is the virtual
photon momentum and $p$ is the nucleon momentum. 
It is known that the structure functions are almost independent
of $Q^2$, which is referred to as Bjorken scaling. 
It indicates that the photon scatters on structureless objects,
which are called partons. They are now identified with quarks and gluons.
Because the variable $x$ is the light-cone momentum fraction carried
by the struck quark, it is possible to extract quark-momentum distributions
from the $F_2$ data.

The $x$ dependence of the parton distributions are expressed by several
parameters at fixed $Q^2$ (=$Q_0^{\, 2}$): for example 
$x\, f_i (x,Q_0^{\, 2})\, =\, A_i \, \eta_i\, x^{\alpha_i}\,
         (1 - x)^{\beta_i}\, (1 + \gamma_i\, x + \rho_i\, \sqrt {x} )$,
and the parameters are optimized so as to explain the existing data.
There are abundant experimental data on unpolarized parton distributions
such as those taken by electron/muon/neutrino deep inelastic scattering,
Drell-Yan, and direct-photon process. With these experimental
data, accurate parton distributions are now known although there are
still detailed issues on the distributions at large $x$ and very small $x$.  
The updated information on the optimum parton distributions is given at
http://durpdg.dur.ac.uk/HEPDATA/HEPDATA.html and
http://www.phys.psu.edu/$\tilde{\ }$cteq.
The major parametrization groups are 
CTEQ (Coordinated Theoretical/Experimental Project on QCD 
      Phenomenology and Tests of the Standard Model),
GRV (Gl\"uck, Reya, and Vogt), and
MRS (Martin, Roberts, and Stirling).
In this section, we introduce how the parton distributions
are extracted from various experimental data for a novice.

\subsection{Valence-quark distributions}\label{VALENCE}

The parton distributions are usually separated into valence-quark,
sea-quark, and gluon distributions.
The valence-quark distribution is defined by the difference
between quark and antiquark distributions: $q_v \equiv q - \bar q$.
It can be obtained by analyzing neutrino deep inelastic data. 
The cross section of the neutrino reaction
${\nu }_{\mu } + p \rightarrow {\mu }^{-} + X$
is expressed as $d\sigma \propto \ell^{\mu\nu} W_{\mu\nu}$
with the lepton tensor $\ell^{\mu\nu}$ and 
hadron tensor $W_{\mu\nu}$ \cite{RGR}.
The lepton tensor is calculated as
\begin{align}
{\ell}^{\mu \nu } = & {\overline {\sum_{spins}}} \, 
\overline{u} (k) {\gamma }^{\mu } (1-{\gamma }_{5}) 
u(k') \cdot \rm \overline{u}(k') {\gamma }^{\nu }
(1-{\gamma }_{5}) u(k)
\nonumber \\
= & \, 2 \left[{\, {k}^{\mu }{k'}^{\nu }
+{k}^{\nu }{k'}^{\mu } - {g}^{\mu \nu }k\cdot k'
+ i{\varepsilon }^{\mu \nu \rho \sigma }
{k}_{\rho }{k'}_{\sigma }\, }\right]\
\ .
\end{align}
Because of the axial vector term $\gamma^\mu \gamma_5$
in the neutrino reaction, there appears the antisymmetric term 
${\varepsilon }^{\mu \nu \rho \sigma}$, which does not exist
in the electron scattering.
Due to the existence of this term in the lepton tensor,
the hadron tensor could also have an antisymmetric one:
\begin{equation}
\! \! \! \!
{W}_{\mu \nu } = - \, W_1 \, ({g}_{\mu \nu }
                   -\frac{{q}_{\mu }{q}_{\nu }}{{q}^{2}})
 + \, W_2 \, \frac{1}{{M}^{2}}\, ({p}^{\mu }
  -\frac{p\cdot q}{{q}^{2}}{q}^{\mu }) ({p}^{\nu }
  -\frac{p\cdot q}{{q}^{2}}{q}^{\nu })
 -  \frac{i}{M} \, W_3 \, {\varepsilon }_{\mu \nu \rho \sigma }
    \, {p}^{\rho } \, {q}^{\sigma}
\, .
\end{equation}
From these expressions, the cross section becomes
\begin{equation}
\! \! \! \!
\frac{{d\sigma }^{\pm }}{d{E}^{'} d\Omega } =
            \frac{{G}_{F}^{2}\, {E'}^{2}}{2  {\pi }^{2} 
            (1+{Q}^{2}/{M}_{W}^{2}{)}^{2}}
  {\bigg [} 
 2 {W}_{1}(\nu,{q}^{2}) \, {\sin}^{2} \frac{\theta}{\rm 2}  
 + {W}_{2}(\nu \rm ,{q}^{2}) \, {\cos}^{2} \frac{\theta}{2}
 \mp  \frac{E+E'}{M} {W}_{3} \,
   {\sin}^{2} \frac{\theta}{2}   {\bigg ]}
\, ,
\end{equation}
where $\pm$ indicates W$^\pm$ exchange processes.
The new term with $W_3$ is associated with the parity violation,
and it is given by the difference between the left and right
cross sections for the W:
${W}_{3}\ \propto {\sigma }_{_L} - {\sigma }_{_R}$.
Instead of $W_3$, the scaling function ${F}_{3} = \nu {\rm W}_{\rm 3}$
is usually used. Because of the ${\sigma }_{_L} - {\sigma }_{_R}$
property, the structure function $F_3$ is given as the quark $-$
antiquark distribution form:
${F}_{3}^{\nu \rm p} = 2 \, (d + s - \overline{u} - \overline{c})$
and
${F}_{3}^{\overline{\nu }p} =
   2 \, (u + c - \overline{d} - \overline{s})$.
Hence, we have
\begin{equation}
{F}_{3}^{\ p} = \frac{{F}_{3}^{\nu \rm p}
+{F}_{3}^{\overline{\nu }p}}{2} = {u}_{v} + {d}_{v}
\ ,
\end{equation}
if $s=\overline{s}$ and $c=\overline{c}$. It is known theoretically,
for example, that the $s$ distribution is not exactly equal to
the $\bar s$. However, because the corrections are expected to be
small, the valence quark distribution $u_v+d_v$ can be
extracted from the neutrino data.

\subsection{Sea-quark distributions}\label{SEA}

The $F_2$ structure function can be used for extracting both valence and
sea quark distributions. It is given by quark distributions weighted by
the square of their charges:
\begin{equation}
F_2(x,Q^2)  =  \sum_i e_i^{\, 2} \, x \, 
                [ \, q(x,Q^2) \, + \, \bar q (x,Q^2) \, ]
\ ,
\end{equation}
in the leading order.
Because it is dominated by the valence distributions at medium and
large $x$ regions, $q_{v}$ can be obtained directly from the $F_2$ data
in these regions.
The $F_2$ structure function at small $x$ is , on the other hand, dominated
by the sea-quark distributions. Therefore, $q_{sea}$ ($=\bar q$) can be obtained
in the small $x$ region from the $F_2$ data. 

The other method is to use Drell-Yan processes
${p}_{1} + {p}_{2} \rightarrow {\mu }^{+}{\mu }^{\bf -} + X$.
Its cross section is described by the $q\bar q$ annihilation processes:
\begin{equation}
s\, \frac{d\sigma}{d\sqrt {\tau }\,  dy} =
   \frac{8\, \pi \, {\alpha }^{2} }{9\, \sqrt {\tau }}
\, \sum\limits_{\rm i}
{e}_{i}^{\, 2}\, 
[\, q({x}_{1},{Q}^{2}) \, \overline{q}({x}_{2},{Q}^{2})
+ \overline{q}({x}_{1},{Q}^{2}) \, q({x}_{2},{Q}^{2})\, ]
\ ,
\end{equation}
where $Q^2$ is the dimuon mass squared: $Q^2=m_{\mu\mu}^2$, 
and $\tau$ is given by $\tau=m_{\mu\mu}^2/s=x_1 x_2$.
The rapidity $y$ is defined by dimuon longitudinal momentum
$P_L^*$ and dimuon energy $E^*$ in the c.m. system:
$y=(1/2)ln[(E^*+P_L^*)/(E^*-P_L^*)]$.
The momentum fractions $x_1$ and $x_2$ can be written 
by these kinematical variables:
$x_1=\sqrt{\tau} e^y$ and $x_2=\sqrt{\tau} e^{-y}$.
In the large $x_{_F}=x_1 - x_2$ region, namely at large $x_1$,
the cross section is dominated by the projectile-quark annihilation
process with an target antiquark:
$d\sigma \propto {q}_{v}({x}_{1})\overline{q}({x}_{2})$
because of $\bar q(x_1) << 1$.
This equation indicates that the sea-quark distribution 
$\overline{q}({x}_{2})$ can be measured if the valence-quark
distributions in the projectile are known.

In these days, the details are investigated for the sea-quark
distributions. Neutrino-induced dimuon events are interpreted
by the charm-production process:
$\nu_\mu +(s,d)\rightarrow$$\mu^- + c$ 
then $c\rightarrow s + \mu^+ + \nu_\mu$,
so that the strange quark distribution can be extracted from
the data. It is now known $s+\bar s\approx 0.5 \, (\bar u+\bar d)$
at $Q^2\approx 20$ GeV$^2$.
Furthermore, flavor asymmetry in light antiquark distributions
was also found by the Gottfried-sum-rule violation and by
the p-n asymmetry in the Drell-Yan experiments \cite{SK}.

\subsection{Gluon distribution}\label{GLUE}

There are two major methods for finding the gluon distribution.
The first one is to use scaling violation data of the $F_2$
structure function, and the second is to use direct photon data.
We mentioned that the structure functions are almost independent of $Q^2$.
However, it is known that they vary as a function of $ln \, Q^2$ even
though it is small. This is called scaling violation.
The $Q^2$ dependence is described by the DGLAP evolution
equations, which are the coupled integrodifferential equations:
\beqn
\frac{\partial}{\partial\ln Q^2} \,
\left(\begin{array}{c} 
  {q}_s (x,Q^2) \\ 
   g (x,Q^2)  
\end{array} \right) =
\frac{\alpha_s (Q^2)}{2\pi} \,
\left( \begin{array}{cc} 
   P_{qq}(x,Q^2) &  P_{qg}(x,Q^2) \\  
   P_{gq}(x,Q^2) &  P_{gg}(x,Q^2) \\ 
\end{array} \right) \otimes
\left( \begin{array}{c}
  {q}_s (x,Q^2) \\
   g(x,Q^2)
\end{array} \right) 
\ ,
\label{scale}
\eeqn
where $q_s= \sum_i (q_i +\bar q_i)$ is the singlet quark distribution. 
The convolution integral $\otimes$ is defined by
$f (x) \otimes g (x) 
   = \int^{1}_{x} f (x/y) \, g(y) \, dy/y$.
Each term describes the process that a parton $p_j$ 
with the nucleon's momentum fraction $y$ 
splits into a parton $p_i$ with the momentum fraction $x$
and another parton. The splitting function $P_{p_i p_j}(z)$
determines the probability that such a splitting process occurs
and the $p_j$-parton momentum is reduced by the fraction $z$. 
The $Q^2$ dependence of the quark distribution is
affected by the gluon distribution through the splitting function.
In the small $x$ region, the right-hand sides of the evolution
equations are dominated by the gluon terms.
Therefore, the gluon distribution can be extracted from
the $Q^2$ dependent data of $F_2$. To be explicit, the $Q^2$ dependence
of $F_2$ is approximately proportional to the gluon distribution:
$\partial F_2 (x,Q^2) / \partial (\ln\, Q^2)
 \approx 10 \, \alpha_s G(2x,Q^2) / (27 \, \pi)$.

\vfill\eject
\begin{wrapfigure}{r}{0.46\textwidth}
   \begin{center}
   \epsfig{file=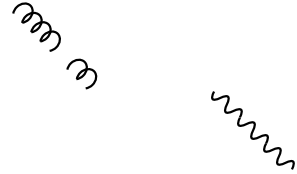,width=5.0cm}
   \end{center}
   \vspace{-0.6cm}
       \caption{\footnotesize QCD Compton process.} 
       \label{fig:photon}
\end{wrapfigure}
The second method is to use direct photon data.
The leading-order cross section is calculated by
the QCD Compton scattering ($gq\rightarrow \gamma q$)
and annihilation ($\bar q q\rightarrow \gamma g$) processes.
Because the gluon distribution is involved in the Compton process
in Fig. \ref{fig:photon}, it is possible to extract it from the
experimental data. The quark and antiquark distributions are also
involved in the reaction; however, they are relatively well known
from deep inelastic scattering data. It is also possible to isolate
the annihilation contribution by using both $pp$ and $p\bar p$ reaction data.
From the scaling violation and direct photon data, the gluon distribution
in the nucleon is now well known except for the small and large $x$
regions. The large $x$ distribution is particularly important for 
judging whether the CDF jet production data could suggest ``subquark"
structure. It is desirable to measure the gluon distribution 
in the $x$ region, $x\sim 0.5$, directly by some kind of experiments.
On the other hand, the small $x$ distribution is not well fixed as
it is obvious from the R1, R2, R3, and R4 type distributions in the MRS
parametrizations \cite{MRS-R}. It is partly because a wide range of $Q^2$ data
is not available in the structure function $F_2$.
These $x$ regions of the gluon distribution should be
studied experimentally in future.

\section{{\bf Activities of the RHIC-Spin-J working group}}\label{PARAMET}
\setcounter{equation}{0}

With the completion of the RHIC facility in mind, we created a working
group within Japan for studying possible spin physics at RHIC.
We have a monthly meeting at Riken among related theorists and
experimentalists. One of the activities is to study the parametrization
of the polarized parton distributions. Even though the study is still
preliminary, we discuss an outline of our efforts.

The parametrization of the unpolarized distributions has been
studied extensively as discussed in the previous section.
On the other hand, the polarized distributions are not well known
yet due to lack of a variety of experimental data.
However, several sets of longitudinally polarized distributions are
proposed by analyzing the $g_1$ structure functions for the proton,
deuteron, and $^3 He$ \cite{PARA}.
With stimulation of these studies and recent polarized measurements,
we initiated a parametrization project within Japan \cite{RHIC-SPIN-J}.
Because this is a part of the Japanese RHIC-Spin activities,
the ultimate purpose is to use our studies for the RHIC-Spin experiments.
At this stage, the group consists of 
Y. Goto, N. Hayashi, M. Hirai, H. Horikawa, S. Kumano,  M. Miyama, T. Morii,
N. Saito, T.-A. Shibata, E. Taniguchi, and T. Yamanishi.
For the time being, this collaboration intends to obtain optimum polarized
parton distributions for explaining all the available experimental data.
We have been working for a year, and we should be able to complete 
our first work in a few months.
It includes complete NLO analyses of the $g_1$
structure functions. The obtained distributions will be used for
experimental studies at RHIC. Furthermore, once new experimental results
are obtained at RHIC or at other facilities,
we try to reanalyze the data for getting updated distributions.

We work in three subgroups: data analysis, parametrization,
and $Q^2$ evolution. The data analysis subgroup (Y. Goto, N. Hayashi,
N. Saito, T.-A. Shibata, and E. Taniguchi) collects all the available
experimental data and investigate their systematic uncertainties. 
At this stage, the data come from measurements of the $g_1$ structure
functions for the proton and ``neutron". The parametrization group 
(T. Morii, H. Horikawa, and T. Yamanishi) tries to understand the meaning
of obtained parameter values. They also spend much time for running
the fitting program. The third subgroup (M. Hirai, S. Kumano, and M. Miyama)
contributes to the $Q^2$ evolution program. The evolution equations are
coupled integrodifferential equations with complicated splitting functions.
This subgroup develops an efficient program for numerical solution
of the evolution equations. 
Our parametrization studies were reported by N. Saito at the spring JPS
meeting in 1997 \cite{JPS}.  Recent studies were also reported by M. Hirai
and T. Yamanishi at this meeting \cite{JPS}.

The distributions are set up at certain $Q^2$ ($\equiv Q_0^{\, 2}$).
Experimental data are taken, in general, at different $Q^2$ points.
The DGLAP evolution equations are used for calculating 
the distribution variations from $Q_0^{\, 2}$ to experimental $Q^2$.
The DGLAP equations assume perturbative QCD, so that both
$Q_0^{\, 2}$ and $Q^2$ have to be in the perturbative region.
The $Q_0^{\, 2}$ is taken typically in the 1$-$4 GeV$^2$ region
except for the GRV distributions, where small $Q_0^{\, 2} \, (\sim$0.3 GeV$^2 )$
is chosen. If it is possible, it is desirable to take very large $Q_0^{\, 2}$
where perturbative QCD is certainly valid. However, because many data
are taken in the $Q^2$ region 1 GeV$^2<Q^2<$a few dozen GeV$^2$, 
we end up taking relatively small $Q_0^{\, 2}$ at this stage.

We provide polarized parton distributions with a number of parameters
at $Q_0^{\, 2}$, for example at 1 GeV$^2$. 
These parameters are determined by fitting the existing experimental
data on $A_1$, which is given by
\beqn
A_1 \cong \frac{g_1(x,Q^2)}{F_1(x,Q^2)} 
  = g_1(x,Q^2) \, \frac{2\, x\, (1+R)}{F_2(x,Q^2)} \ ,
\label{geta}
\eeqn
where the $F_1$ structure function is related to the
$F_2$ structure function and the $R$ function by
$R=(F_2-2xF_1)/(2xF_1)$.
The $g_1$ structure function is calculated by the convolution integral of
the polarized parton distributions with the coefficient functions:
\begin{align}
{g}_{1}(x,{Q}^{2}) = \frac{1}{2}\, \sum\limits_q e_q^2 \, 
      \bigg\{ & \int_{x}^{1} \frac{dy}{y}\, \Delta C_q (x/y,Q^2)\, 
      [\, \Delta q(y,Q^2)\, +\, \Delta \overline{q}(y,Q^2)\, ]
\nonumber \\
  + & \int_{x}^{1} \frac{dy}{y}\, \Delta C_g (x/y,Q^2)\, \Delta g(y,Q^2)\, 
        \bigg\}
\ .
\end{align}
It is not a good idea to analyze the published $g_1$
values directly because it is often assumed in the experimental 
analyses that $A_1$ is $Q^2$ independent. As it is obvious from
the results in the next section (see Figs. \ref{fig:a-1} and \ref{fig:a-3}),
it is not a correct assumption. However, considering the present
experimental errors, one cannot help using the assumption.
In our analysis, the asymmetry $A_1$ data are analyzed without
using the $Q^2$ independent assumption.
We investigated various functional forms of the initial distributions;
however, the studies have not been completed yet. 
For example, the $x$ dependence is given as
$x\, \Delta f_i (x,Q_0^{\, 2})\, =\, A_i \, \eta_i\, x^{\alpha_i}\,
         (1 - x)^{\beta_i}\, (1 + \gamma_i\, x + \rho_i\, \sqrt {x} )$
in the analysis of Gehrmann and Stirling (GS). 
The $F_2$ structure function is given in the similar way as
\vfill\eject
\begin{align}
F_2(x,Q^2) = \sum\limits_q e_q^2 \, x \, 
    \bigg\{ & \int_x^1 \frac{dy}{y}\, C_q (x/y,Q^2)\, 
    [\, q(y,Q^2) + \overline{q}(y,Q^2)\, ]
\nonumber \\
+ & \int_x^1 \frac{dy}{y}\, C_g(x/y,Q^2)\, g(y,Q^2)\, \bigg\}
\ .
\end{align}
\begin{wrapfigure}{r}{0.46\textwidth}
  \vspace{-0.3cm}
   \begin{center}
   \epsfig{file=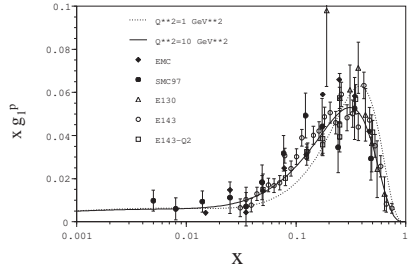,width=6.0cm}
   \end{center}
   \vspace{-1.1cm}
       \caption{\footnotesize Fitting results (preliminary)
           {\normalsize \cite{JPS}}.}
       \label{fig:g1p}
\end{wrapfigure}
Because the unpolarized distributions are not studied by our group
at least at this stage , we take distributions
$f_i (x,Q_0^{\, 2})$ by CTEQ, GRV, or MRS.
With these preparations, we are ready for the optimization.
We compare the theoretical asymmetries $A_1$ with the experimental
data and minimize
\begin{equation}
\chi^2\, =\, \sum\limits 
    \frac{ (A_1^{data}\, -\, A_1^{calc})^2}
         { (\sigma _{A_1^{data}})^2}
\end{equation}
by using the CERN subroutine MINUIT.
Because the only information comes from the longitudinally polarized
structure function $g_1$, the polarized sea-quark and gluon distributions
are not well determined at this stage.
We tried to find the optimum parameter set by fitting the data
with the $Q^2$ evolution effects.
In Fig. \ref{fig:g1p}, our preliminary fitting results are shown
with various experimental data for $g_1^{\, p}$ \cite{JPS}.
The dotted curve is the initial distribution at $Q^2$=1 GeV$^2$,
and the solid one is the distribution at $Q^2$=10 GeV$^2$.
We are still working on obtaining the best fit.

We have been studying the parametrization for almost a year.
The leading order (LO) fitting program was completed a half year ago,
and we try to understand physics meaning of the obtained parameters.
On the other hand, we completed the NLO program and the NLO analysis
is in progress. 
We should be able to report our results within a few months.

\section{{\bf Numerical solution of longitudinal and transversity
              $Q^2$ evolution equations}}\label{EVOL}
\setcounter{equation}{0}

The DGLAP evolution equations are frequently used in theoretical
and experimental analyses. In fact, our evolution program is used
in the parametrization studies in the last section.
The evolution equations are coupled integrodifferential
equations with complex splitting functions in the NLO case.
It is, therefore, important to create a computer program to solve them
accurately and to publish it so that other people could use the subroutine.
We have been working on this topic for several years
\cite{T-EVOL,L-EVOL,UNPOL}.
In this section, we report the numerical solution of
the $Q^2$ evolution equations for the
longitudinally polarized and transversity distributions.

The longitudinal DGLAP equations are given
in the similar way with the unpolarized ones. The nonsinglet equation is
\vfill\eject
\beqn
\frac{\partial}{\partial\ln Q^2} \, \Delta \widetilde q_{_{NS}}(x,Q^2)\ = 
\frac{\alpha_s (Q^2)}{2\pi} \,  
\Delta \widetilde P_{q^\pm, {NS}} (x) \otimes \Delta 
\widetilde q_{_{NS}}(x,Q^2) 
\ ,
\label{nonap}
\eeqn
where $\Delta \widetilde q_{_{NS}} = x \Delta q_{_{NS}}$ is 
a longitudinally polarized nonsinglet parton distribution with
$\Delta q = q_{\uparrow}-q_{\downarrow}$.
The parton distribution and the splitting function multiplied by
$x$ are denoted as $\widetilde f (x) = x f(x)$.
The function $\Delta \widetilde P_{q^\pm, {NS}} = x \Delta P_{q^\pm, {NS}}$ 
is the polarized nonsinglet splitting function.
The notation $q^\pm$ in the splitting function indicates
a $\Delta q^+ = \Delta q + \Delta \bar q$ or 
  $\Delta q^- = \Delta q - \Delta \bar q$ distribution type.
The singlet $Q^2$ evolution is described by the coupled integrodifferential
equations due to the gluon participation:
\beqn
\frac{\partial}{\partial\ln Q^2} 
\left(\begin{array}{c} 
  \Delta \widetilde{q}_s (x,Q^2) \\ 
  \Delta \widetilde g (x,Q^2)  
\end{array} \right) =
\frac{\alpha_s (Q^2)}{2\pi} 
\left( \begin{array}{cc} 
  \Delta \widetilde P_{qq}(x,Q^2) & \Delta \widetilde P_{qg}(x,Q^2) \\  
  \Delta \widetilde P_{gq}(x,Q^2) & \Delta \widetilde P_{gg}(x,Q^2) \\ 
\end{array} \right) \otimes
\left( \begin{array}{c}
  \Delta \widetilde{q}_s (x,Q^2) \\
  \Delta \widetilde g(x,Q^2)
\end{array} \right) 
\! .
\label{singlet}
\eeqn
The singlet quark distribution is defined by 
$\Delta \widetilde{q}_s \equiv \sum_i^{N_{f}} x\, \Delta q_i^+ $ 
where $i$ is the flavor, and $\Delta \widetilde g(x,Q^2) $ is the gluon
distribution $\Delta \widetilde g = x \, (g_{+1}-g_{-1})$.
The LO and NLO \cite{G1NLO} evolution equations are described by
the above integrodifferential equations. The differences between
the LO and NLO are contained in the running coupling constant $\alpha_s$
and in the splitting functions.
The renormalization scheme is the modified minimal subtraction
scheme ($\overline{MS}$) in the NLO analysis.

In addition to the unpolarized and longitudinally polarized evolution
equations, the NLO evolution for the transversity distribution was
completed recently \cite{H1NLO}.
The transversity distribution is denoted as $h_1$, $\delta q$, or
$\Delta_{_T} q$. Throughout this paper, the notation $\Delta_{_T} q$
is used. Because the gluon does not couple due to the chiral-odd nature of 
$\Delta_{_T} q$, the evolution is described by a single integrodifferential
equation,
\begin{equation}
\frac{\partial}{\partial\ln Q^2} \, \Delta_{_T} q^\pm (x,Q^2) \, = 
\frac{\alpha_s (Q^2)}{2\pi} \,  
\Delta_{_T} P_{q^\pm} (x) \otimes \Delta_{_T} q^\pm (x,Q^2) 
\ .
\label{eqn:DGLAP1}
\end{equation}
Numerical analysis of the transversity evolution is simple in
the sense that it is given only by the ``nonsinglet" type equation. 

The longitudinally polarized and transversity distributions
will be measured at RHIC by polarized p+p reactions.
However, they are measured at different $Q^2$ points.
In order to extract information on the polarized parton distributions,
an accurate program for solving the evolution equations should be developed.
We have been studying on this topic, and
the following discussions are based on the numerical results
in Refs. \cite{T-EVOL,L-EVOL}.
We simply divide the variables $Q^2$ and $x$ into small steps.
Then, the integration and differentiation are defined by
\begin{align}
\frac{d f(x)}{dx} &= \frac{f(x_{m+1})-f(x_m)}{\delta x_m} \ ,\\
\int f(x)\ dx &= \sum_{m=1}^{N_x} \delta x_m \, f(x_m) \ .
\end{align}
With these replacements, the evolution equations could be solved rather
easily. We call this method a ``brute-force" method \cite{L-EVOL}.
The better method, by means of convergence and computing time, was
studied in Ref. \cite{T-EVOL} by changing the $x$ integration 
for the Simpson's method.
As the step number becomes larger, the numerical accuracy becomes
better. We obtain the numerical accuracy better than 1\% in
the $x$ region $10^{-8}<x<0.8$ with two hundred $Q^2$ steps
and one thousand $x$ steps in the brute-force method
and with fifty $Q^2$ steps and five hundred $x$ steps
in the Simpson's method. 

\begin{wrapfigure}{r}{0.46\textwidth}
  \vspace{-0.3cm}
   \begin{center}
   \epsfig{file=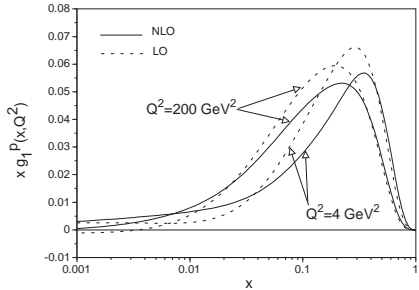,width=6.0cm}
   \end{center}
   \vspace{-0.6cm}
       \caption{\footnotesize  
           $Q^2$ evolution results for $g_1^p$
           {\normalsize \cite{L-EVOL}}.}
       \label{fig:xg1}
\end{wrapfigure}
We show evolution results of the longitudinally polarized
parton distributions and transversity distributions by using
the developed programs, BFP1 \cite{L-EVOL} and H1EVOL \cite{T-EVOL}.
First, the longitudinal evolution is discussed.
The initial distributions are the GS set-A distributions at $Q^2$=4 GeV$^2$. 
The evolution results are shown in Fig. \ref{fig:xg1}.
The same GS-A distributions are assumed for the LO and NLO cases,
so that the differences between the LO and NLO curves at $Q^2$=4 GeV$^2$
are solely due to the NLO coefficient
functions. The $xg_1$ distributions are evolved to those at 
$Q^2$=200 GeV$^2$. They increase at small $x$ and decrease at large $x$.
The major differences between the LO and NLO distributions at 
$Q^2$=200 GeV$^2$ are also due to the coefficient functions.

\vspace{-0.6cm}
\noindent
\begin{figure}[h]
\parbox[t]{0.46\textwidth}{
   \begin{center}
       \epsfig{file=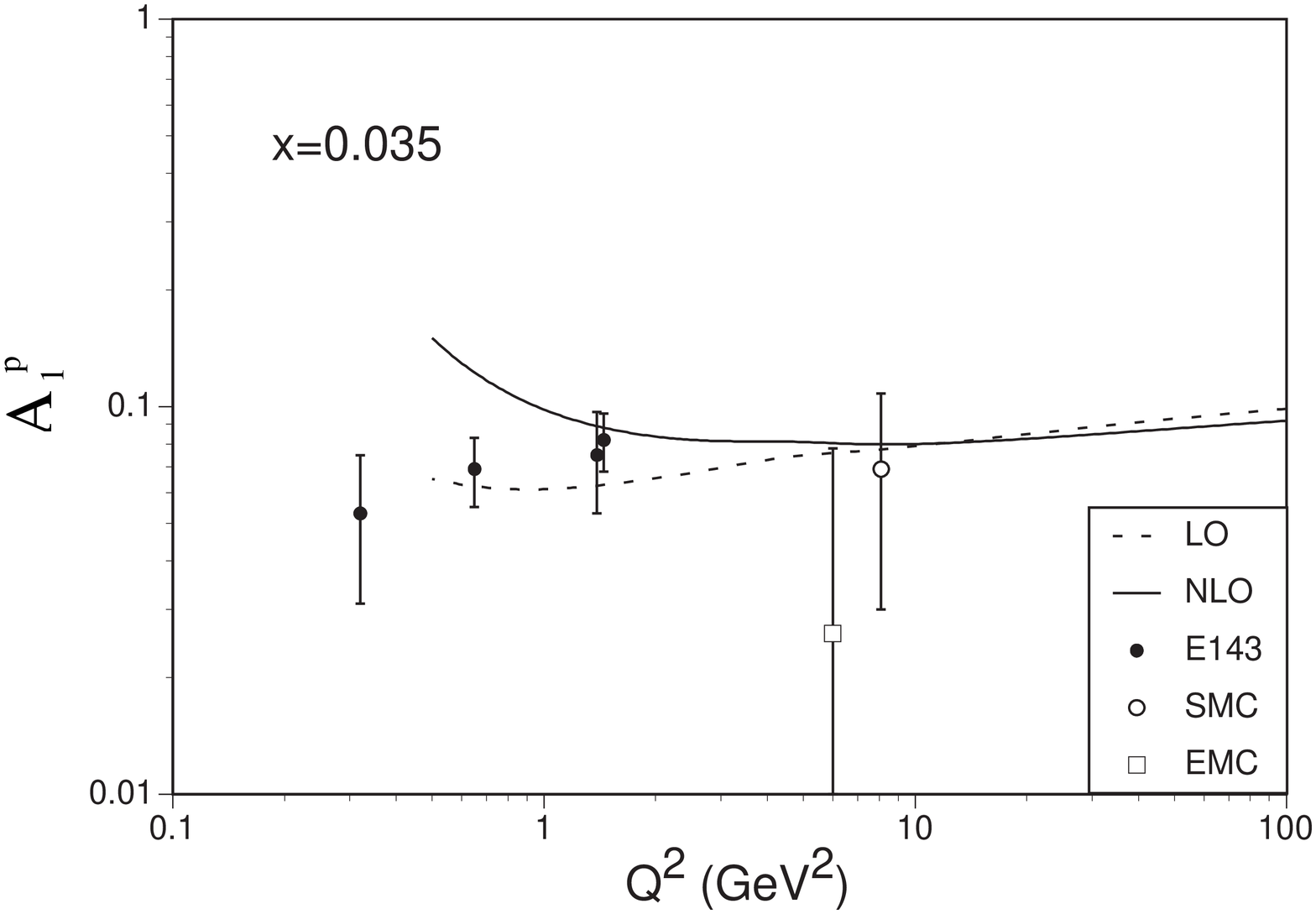,width=6.0cm}
   \end{center}
   \vspace{-0.8cm}
       \caption{\footnotesize
                $Q^2$ dependence of $A_1^p$
                is calculated in the LO (dashed curve) 
                and NLO (solid curve) cases at $x=0.035$
                {\normalsize \cite{L-EVOL}}.}
       \label{fig:a-1}
}\hfill
\parbox[t]{0.46\textwidth}{
   \begin{center}
   \epsfig{file=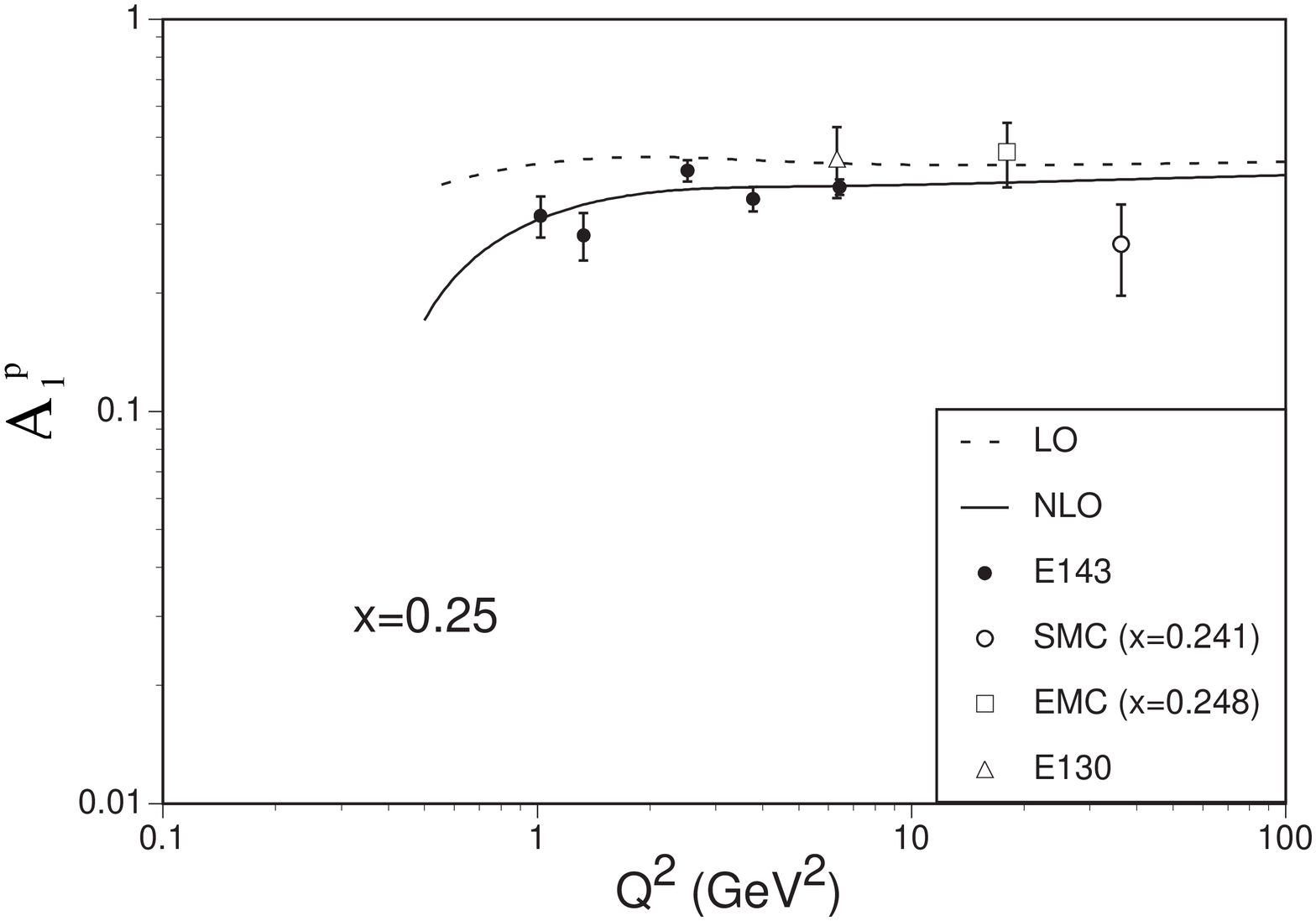,width=6.0cm}
   \end{center}
   \vspace{-0.8cm}
       \caption{\footnotesize Spin asymmetry $A_1^p$ at $x=0.25$
               {\normalsize \cite{L-EVOL}}.}
       \label{fig:a-3}
}
\end{figure}

In Fig. \ref{fig:a-1}, our evolution curves at $x$=0.035 are shown
with the asymmetry $A_1$ data. 
The dashed and solid curves indicate the LO and NLO evolution 
results, respectively. In the large $Q^2$ region, both results are almost
the same; however, they differ significantly at small $Q^2$ particularly
in the region $Q^2<\, $2 GeV$^2$.
In the medium $x$ region ($x$=0.25), the difference
becomes larger again at small $Q^2$ as shown in Fig. \ref{fig:a-3}.
From these figures, we find that the asymmetry has $Q^2$ dependence
although it is not large. 
People used to assume that the asymmetry
is independent of $Q^2$ by neglecting the $Q^2$ evolution difference
between $g_1$ and $F_1$ in analyzing the experimental data.
We find clearly that it is not the case. For a precise analysis,
the $Q^2$ dependence in the asymmetry has to be taken into account.

Next, we discuss the transversity evolution $\Delta_{_T} q$.
It has a chiral-odd property so that it cannot be found
in inclusive deep inelastic electron scattering.
It is expected to be measured in the RHIC-Spin project
by the transversely polarized Drell-Yan processes.
The distribution $\Delta_{_T} q$ is 
the probability to find a quark with spin polarized along
the transverse spin of a polarized proton minus
the probability to find it polarized oppositely.
There is a problem in studying the transversity evolution
in the sense that the input distribution is not available.
However, it is known within quark models that the transversity
distributions are almost the same as the corresponding
longitudinally polarized distributions. Therefore, we may
use a longitudinal distribution as an input transversity
distribution at small $Q^2$, where the quark models would be valid.
This consideration is fine at $Q^2\sim \Lambda_{QCD}^{\, 2}$.
On the other hand, the perturbative QCD cannot be used
at such a small $Q^2$. In the following numerical analysis,
we simply assume $\Delta q=\Delta_{_T} q$ at relatively large $Q^2$
(=$\, 4$ GeV$^2$) by taking the GS-A distributions as the input.

\begin{figure}[h]
\parbox[t]{0.46\textwidth}{
   \begin{center}
       \epsfig{file=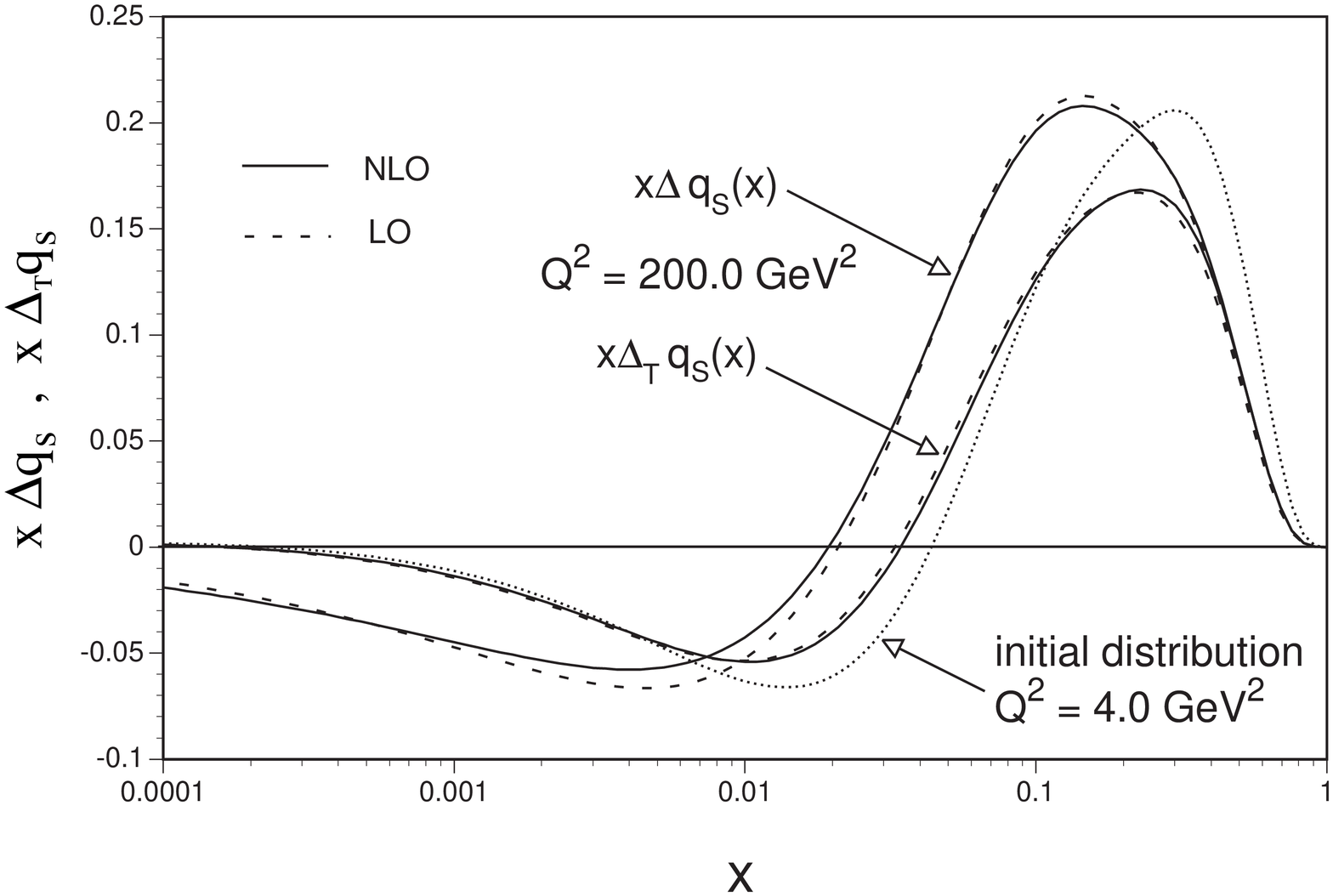,width=6.0cm}
   \end{center}
   \vspace{-0.6cm}
       \caption{\footnotesize
The singlet transversity evolution results
are compared with those of longitudinally polarized singlet distribution
{\normalsize \cite{T-EVOL}}.}
       \label{fig:xqts}
}\hfill
\parbox[t]{0.46\textwidth}{
   \begin{center}
   \epsfig{file=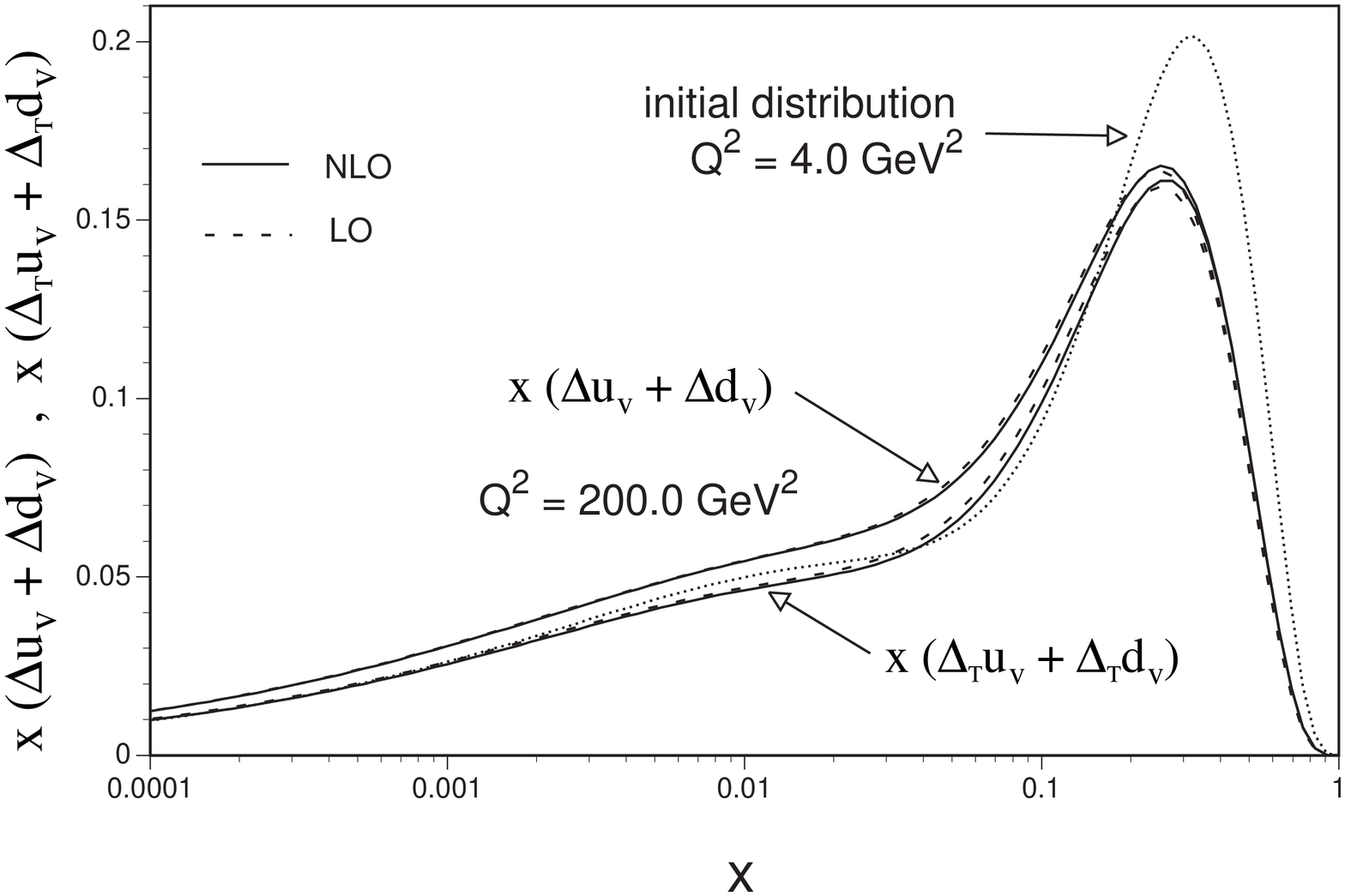,width=6.0cm}
   \end{center}
   \vspace{-0.6cm}
       \caption{\footnotesize 
Evolution results of the valence-quark transversity distribution
$x(\Delta_{_T}u_v+\Delta_{_T}d_v)$ are compared with those of
the longitudinally polarized one
{\normalsize \cite{T-EVOL}}.}
       \label{fig:xqtv}
}
\end{figure}

We show the evolution results of two distributions,
the singlet distribution $x\sum_i (\Delta_{_T} q +\Delta_{_T} \bar q)$
and the valence distribution $x (\Delta_{_T} u_v+\Delta_{_T} d_v)$.
First, the singlet evolution is shown in Fig. \ref{fig:xqts}.
The same GS-A distribution is assumed for the transversity and 
longitudinally-polarized parton distributions at $Q^2$=4 GeV$^2$.
The initial distribution is shown by the dotted curve.
It is evolved to the distributions at $Q^2$=200 GeV$^2$ by the 
transverse or longitudinal evolution equation.
The transversity NLO effects increase the evolved distribution
at medium-large $x$ and also at small $x$ ($<$0.01), 
and they decrease the distribution in the intermediate
$x$ region ($0.01<x<0.1$). 
The transversity evolution differs significantly from
the longitudinal one either in the LO case or in the NLO.
The evolved transversity distribution $\Delta_{_T} q_s$
is significantly smaller than the longitudinal one $\Delta q_s$
in the region $x\sim 0.1$. The magnitude of $\Delta_{_T} q_s$
itself is also smaller than that of $\Delta q_s$ at very small $x$
($<0.07$). 

Second, the evolution of the valence-quark distribution is shown
in Fig. \ref{fig:xqtv}.
The evolution calculations are done in the same way.
The dotted curve is the initial distribution, and the evolved
distributions are solid and dashed curves.
The evolved transversity distributions are significantly
smaller than the longitudinally polarized ones, in particular
at small $x$.

\vspace{-0.6cm}
\noindent

We have created the useful computer programs, BFP1 and H1EVOL,
for solving the DGLAP evolution equations for the longitudinally
polarized parton distributions and for the transversity distributions.
They should be convenient for theoretical and experimental researchers,
who are involved in high-energy spin physics. 
Our program can be obtained from M. Hirai, SK, or M. Miyama upon email
request. For the details, the reader may look at the WWW home page:
http://www.cc.saga-u.ac.jp/saga-u/riko/physics/quantum1/program.html.

\vfill\eject
\section{{\bf Summary}}\label{SUM}
\setcounter{equation}{0}

Since the Riken decided to join the RHIC-Spin project, we have been
studying the physics aspects of the project. 
In particular, the parametrization of polarized parton distributions
has been investigated among the members.
For general audience at this symposium, the method of extracting
unpolarized parton distributions is first discussed.
Then, our parametrization project was explained although
the results are still preliminary.
The optimum parton distributions are obtained by fitting existing
experimental data of the spin asymmetry $A_1$.
On the other hand, we have been working on creating
efficient computer programs of solving the $Q^2$ evolution equations.
Now, the evolution programs for the unpolarized,
longitudinally polarized, and transversity parton distributions
are available. Our efforts should be useful for researchers
in the area of high-energy spin physics.

\vspace{0.3cm}
\begin{center}
{\bf Acknowledgments} \\
\end{center}
\vspace{-0.1cm}

SK thanks the members of the RHIC-Spin-J working group
                (Y. Goto, N. Hayashi, M. Hirai, H. Horikawa, 
                 S. Kumano,  M. Miyama, T. Morii, N. Saito, 
                 T.-A. Shibata, E. Taniguchi, and T. Yamanishi)
for discussions. The material in section \ref{PARAMET} is quoted
from their studies \cite{RHIC-SPIN-J,JPS}.
The section \ref{EVOL} is based on the works with
M. Hirai and M. Miyama \cite{T-EVOL,L-EVOL}.

\vspace{0.5cm}
\noindent
{* Email: kumanos@cc.saga-u.ac.jp;}  \\

\vspace{-0.55cm}
{http://www.cc.saga-u.ac.jp/saga-u/riko/physics/quantum1/structure.html.} \\

\vspace{-0.4cm}

\end{document}